\documentclass[aps,prl,reprint, superscriptaddress, bibliography]{revtex4-1}
\usepackage[section]{placeins}

\usepackage{lineno}
\usepackage{graphicx}  
\usepackage[caption=false]{subfig}
\usepackage{dcolumn}   
\usepackage{tabularx}
\usepackage{bm}        
\usepackage{amssymb}   
\usepackage{amsmath,stmaryrd}
\usepackage{dsfont}
\usepackage{booktabs}
\usepackage{blkarray, multirow, graphicx, diagbox, color, colortbl}
\usepackage[dvipsnames]{xcolor}
\usepackage{bbm, bbold}
\usepackage{glossaries}
\usepackage{hyphenat}
\usepackage{ifthen}
\usepackage[colorlinks, linkcolor = blue, citecolor = blue, filecolor = black, urlcolor = blue]{hyperref}
\usepackage{xkeyval}
\usepackage{moreverb}
\usepackage{rotating}
\usepackage{wrapfig}
\usepackage{slashbox}
\usepackage{xspace}
\usepackage{nicefrac}
\usepackage[]{units}
\usepackage{physics}
\usepackage{braket}
\usepackage[inline]{enumitem}
\usepackage{tabto}
\usepackage{listings}
\usepackage{xstring}
\usepackage{pgfplots}
\usepackage[version=4]{mhchem}
\pgfplotsset{compat=1.18}
\def\ReplaceStr#1{%
	\IfSubStr{#1}{p}{%
		\StrSubstitute{#1}{p}{.}}{#1}}

\usepackage[capitalise]{cleveref} %
\usepackage{chemformula}
\usepackage{algorithm}

\newcommand{\ie}{i.e.~}

\newcommand{\sgn}[1]{\text{sign}({#1})}
\newcommand{\suppmat}{Supplemental Material \cite{supp_mat}}

\newcommand{\arrowline}[2]{%
  \mathrel{%
    \begin{tikzpicture}[baseline={(0,0)}]
      \draw[thick, color=#1] (-0.3,0.09) -- (0.3,0.09);
      \ifthenelse{\equal{#2}{up}}{
        \draw[->, thick, >=stealth] (0,-0.1) -- (0,0.3);
      }{
        \draw[->, thick, >=stealth] (0,0.3) -- (0,-0.1);
      }
    \end{tikzpicture}%
  }%
}

\definecolor{colorA}{rgb} {0.9607843137254902, 0.25882352941176473, 0.25882352941176473} 
\definecolor{colorB}{rgb} {0.9607843137254902, 0.6666666666666666, 0.25882352941176473} 
\definecolor{colorC}{rgb} {0.6, 0.6, 0.6} 
\definecolor{colorD}{rgb} {0.6549019607843137, 0.9607843137254902, 0.25882352941176473} 
\definecolor{colorE}{rgb} {0.25882352941176473, 0.9607843137254902, 0.8784313725490196} 
\newacronym{SOC}{SOC}{spin-orbit coupling}
\newacronym{QAHE}{QAHE}{quantum anomalous Hall effect}
\newacronym{QAH}{QAH}{quantum anomalous Hall}
\newacronym{BZ}{BZ}{Brillouin zone}
\newacronym{DFT}{DFT}{density functional theory}
\newacronym{MAE}{MAE}{magnetic anisotropy energy}
\newacronym{WSM}{WSM}{Weyl semimetal}
\newacronym{TB}{TB}{tight-binding}
\newacronym{IR}{irrep}{irreducible representation}
\newacronym{DSG}{DSG}{double space group}
\newacronym{HWCC}{HWCC}{hybrid Wannier charge center}
\newacronym{KS}{KS}{Kohn-Sham}
\newacronym{SG}{SG}{space group}
\newacronym{WF}{WF}{Wannier function}
\newacronym{1D}{1D}{one-dimensional}
\newacronym{2D}{2D}{two-dimensional}
\newacronym{3D}{3D}{three-dimensional}
\newacronym{WHSM}{WHSM}{Weyl half-semimetal}
\newacronym{DOS}{DOS}{density of states}
\newacronym{WP}{WP}{Weyl point}
\newacronym{GGA}{GGA}{generalized gradient approximation}
\newacronym{PAW}{PAW}{projector augmented wave}
\newacronym{DFPT}{DFPT}{density functional perturbation theory}
\graphicspath{{figures/main_text/}}
\begin{document}
\author{Leonard Werner Pingen} 
\thanks{These authors contributed equally to this work}
\affiliation{Trinity College, University of Cambridge, Trinity Street, Cambridge CB2 1TQ, UK}
\affiliation{Centre for Scientific Computing, Cavendish Laboratory, University of Cambridge, J.\,J.\,Thomson Avenue, Cambridge CB3 0US, UK}
\author{Jiaqi Wu} 
\thanks{These authors contributed equally to this work}
\affiliation{Peterhouse, University of Cambridge, Trumpington Street, Cambridge CB2 1RD, UK}
\author{Bo Peng} 
\email{bp432@cam.ac.uk}
\affiliation{Theory of Condensed Matter Group, Cavendish Laboratory, University of Cambridge, J.J.Thomson Avenue, Cambridge CB3 0US, UK}
\def\thetitle{Tunable quantum anomalous Hall effect in fullerene monolayers}
\title{\thetitle}
\begin{abstract}
Nearly four decades after its theoretical prediction, the search for material realizations of \gls{QAHE} remains a highly active field of research.
Many materials have been predicted to exhibit \gls{QAH} physics under feasible conditions but the experimental verification remains widely elusive.
In this work, we propose an alternative approach towards \gls{QAH} materials design by engineering customized molecular building blocks.
We demonstrate this ansatz for a two-dimensional honeycomb lattice of \ce{C26} fullerenes, which exhibits a ferromagnetic ground state and thus breaks time\hyp reversal symmetry.
The molecular system is found to be highly tunable with respect to its magnetic degrees of freedom and applied strain, giving rise to a rich phase diagram with Chern numbers $C=\pm2,\pm1,0$.
Our proposal offers a versatile platform to realize tunable QAH physics under accessible conditions and provides an experimentally-feasible approach for synthesis of \gls{QAH} materials using molecules.
\end{abstract}
\maketitle
The quantization of transverse conductivity in a \gls{2D} electron gas under the effect of a perpendicular magnetic field is known as the quantum Hall effect.
Soon after its experimental observation in 1980\,\cite{Klitzing1980}, the underlying mechanism was explained theoretically\,\cite{Thouless1982} by associating its remarkable robust quantization with a topological invariant called Chern number\,\cite{Niu1985, Kohmoto1985}.
In 1988, Haldane demonstrated that the same effect could be realized on a lattice even in the absence of net magnetic flux\,\cite{Haldane1988}, which is referred to as the \acrfull{QAHE}\,\cite{Qi2011, Hasan2010, Weng2015_2, He2018, Chang2023, Liu2016}.
A material realization of such a bulk\hyp insulating system would offer immense potential from a technological perspective: Possible application range from high\hyp precision quantum sensing devices operated near topological phase transitions\,\cite{Korkusinski2014, Liu2021} to ultra\hyp low power consumption topological transistors\,\cite{Nadeem2024, Gilbert2021} or the so\hyp called Chern networks\,\cite{Gilbert2025} which can be used for information processing by exploiting the emergence dissipationless edge states.

Haldane himself, however, already questioned the feasibility to implement his model, which indeed turned out to be a challenging endeavor.
The first observation of the \gls{QAHE} was reported in 2013\,\cite{Chang2013} at a temperature of $30 ~ \text{mK}$ and thus clearly below ambient conditions, preventing its immediate technological applicability.
It was subsequently demonstrated that the temperatures could be increased for materials with intrinsic magnetic order instead of introducing the required time\hyp reversal symmetry breaking via magnetic dopants\,\cite{Deng2020}.
As another class of material platforms\,\cite{Chang2023}, the \gls{QAHE} has further been realized in moir{\'e} materials at $\lesssim 3~\mathrm{K}$\,\cite{Serlin2020}.
Nevertheless, for all three successfully demonstrated implementations, observation of the \gls{QAHE} at experimentally-feasible temperature remains challenging.
Besides the problem of reconciling ambient, experimental conditions under which such topological phases of matter emerge, it is crucial to identify more tunable materials, which allow for sensitive switching mechanisms between different topological phases.
For early realizations, this could only be achieved by applying large magnetic fields, again hindering practical applicability.
Over the past years, several material platforms have been theoretically and computationally predicted to exhibit flexible \acrfull{QAH} physics\,\cite{Liang2025, Zhang2025, Das2024, Zhang2023, Wu2023, Li2022, Bultinck2020, Ge2020, Zhao2019, Wang2017, Wu2014, Liu2022_2, Yang2025, Zhao2023} with the highest estimated Curie temperature $\sim 550~\mathrm{K}$\,\cite{Chen2021}.
However, precise atomic engineering necessary to synthesize such compounds remains challenging in practice\,\cite{Mannix2017}.
\newline \indent
Here we pursue an alternative approach to realize \gls{QAH} materials using stable molecular building blocks.
We refer to this method as the \textit{superatom ansatz}: The essential idea is to design material properties based on features exhibited by stable molecular units which behave like atoms.
This ansatz allows for increased chemical tunability while facilitating less challenging implementation compared to atomic building blocks\,\cite{Peng2022c,Peng2023,Peng2025,Peng2025c,Wu2025a,Shaikh2025}.
In light of recent progress in synthesizing \gls{2D} fullerenic networks\,\cite{Hou2022, Meirzadeh2023}, we demonstrate this approach in a honeycomb structure of \ce{C26} molecules.
The origin of magnetization in fullerenic materials has recently been rationalized\,\cite{Wu2025,Peng2025d,Wu2025b}.
Here, individual \ce{C26} molecules exhibit non\hyp vanishing magnetization due to a spin\hyp polarized ground state.
Ferromagnetic exchange interactions facilitate time\hyp reversal symmetry breaking, which is mandatory in order to observe \gls{QAH} physics.
The high degree of symmetry facilitated by the underlying hexagonal lattice commonly stabilizes massless quasiparticles of Weyl type, which can open topologically non\hyp trivial band gaps upon considering \gls{SOC}\,\cite{Zhang2025}.
We confirm this behavior for the direct band gap from first\hyp principles.
The weak \gls{SOC} in carbon systems implies a high degree of magnetic anisotropy, enabling topological phase transitions by controlling the orientation of magnetization.
Uniaxial strain provides further degrees of freedom, which can be exploited to induce transitions between distinct topological phases.
Based on theoretical considerations and Wannier \gls{TB} models, we establish the rich phase diagram of honeycomb \ce{C26} polymeric fullerene monolayers.
Upon inducing a global band gap for topological regimes, phase boundaries resemble \gls{2D} \glspl{WHSM}, featuring Weyl\hyp like dispersion in a single spin channel around the Fermi energy\,\cite{You2019}.
The plethora of topological phases in combination with easily-accessible control mechanisms in molecular systems provide an unprecedented material realization for exploring \gls{QAH} physics in practice -- in particular regarding ease of synthesis and external control.
Our results demonstrate the strengths of the superatom ansatz, which is anticipated to present a versatile approach towards designing functional quantum materials.
\textit{Methods-- }
Our numerical results are based on a quasi\hyp \gls{2D} \ce{C26} monolayer structure which is relaxed using the conjugate gradients method\,\cite{Payne1992} implemented in the Vienna \textit{Ab initio} Simulation Package (\textsc{VASP})\,\cite{Kresse1993, Kresse1996_1, Kresse1996_2, Kresse1999} based on \gls{DFT}\cite{Hohenberg1964, Kohn1965} within the \gls{GGA} using the PBEsol functional\,\cite{Perdew2008}.
%
%
%
A plane\hyp wave cutoff of $800 ~ e\text{V}$ is used on a $5 \times 5 \times 1$ Monkhorst\hyp Pack $\mathbf{k}$\hyp point mesh for the 2D \gls{BZ}.
Electrons in 1s orbitals are frozen using the \gls{PAW} method.
A vacuum layer of $\sim 35 ~ \text{\AA}$ is introduced between two monolayers to study the quasi\hyp \gls{2D} system.
Phonon band structures are obtained using the implementation of \gls{DFPT} \cite{Baroni2001,Gonze1997} in \textsc{VASP} for a $2\times2$ supercell and visualized using \textsc{PHONOPY}\,\cite{Togo2023_1, Togo2023_2}.    
To find the magnetic ground state, a \gls{TB} model is constructed using \textsc{Wannier90}\,\cite{Pizzi2020} by projecting onto $\sigma$ bond centres between all carbon atoms and one extra $p_z$ orbital of each $sp^2$ carbon atom, as widely used in fullerene systems\,\cite{Mostofi2008}.
The exchange interactions between fractional spins are calculated using \textsc{TB2J}\,\cite{He2021} and the magnon dispersion is obtained using \textsc{Magnopy}\,\cite{Holstein1940, Colpa1978, Ivanov2021} with $1/9~\mu_\mathrm{B}$ on each magnetic atom. 
\textit{Structural \& magnetic properties-- }
\begin{figure}[tb!]
    \centering
    \subfloat[\label{fig:structure}]{
      \begin{minipage}[t]{\linewidth}
        \centering
        \includegraphics[width=\linewidth]{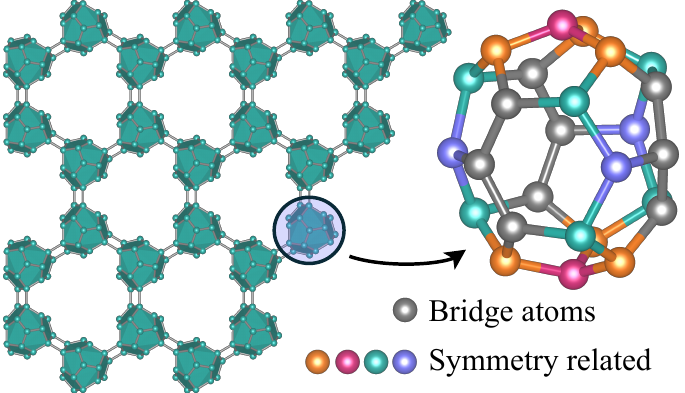}
      \end{minipage}
    }
    \vspace{0em}
    \subfloat[\label{fig:phonons}]{
      \begin{minipage}[t]{0.49\linewidth}
        \centering
        \includegraphics[width=\linewidth]{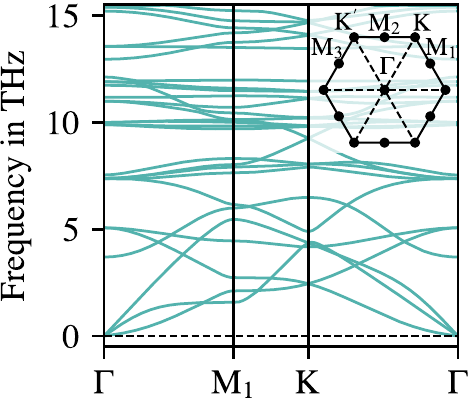}
      \end{minipage}
    }
    \subfloat[\label{fig:magnons}]{
      \begin{minipage}[t]{0.49\linewidth}
        \centering
        \includegraphics[width=\linewidth]{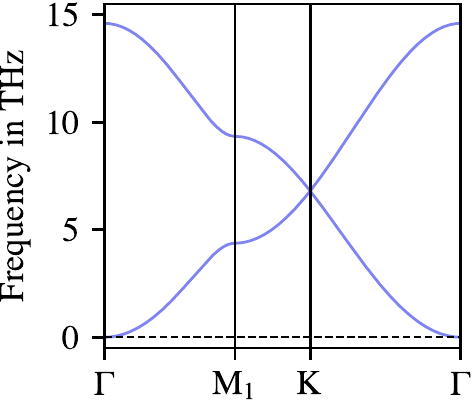}
      \end{minipage}
    }
    \caption{
        (a) Hexagonal lattice of \ce{C26} fullerenes.
        An isolated molecule is shown on the right.
        Here, gray colored carbon atoms form the inter\hyp molecular bonds of the quasi\hyp \gls{2D} structure.
        The colors of the remaining ions indicate the sets of weakly hybridized $p$\hyp orbitals that are closed under the action of the point group.
        (b) Low-frequency phonon dispersion.
        The \gls{BZ} and high\hyp symmetry lattice momenta are shown in the top right.
        (c) Low-frequency magnon dispersion.
    }
    \label{fig:structure_bz}
\end{figure}
We demonstrate the superatom ansatz by employing \ce{C26} molecules as stable building blocks.
The constituting carbon atoms bond to form three hexagons separated by a total of twelve pentagons, comprising molecular symmetry group $D_{3h}$.
\Cref{fig:structure} shows a \gls{2D} structure built from \ce{C26} molecules arranged on a honeycomb lattice.
Its dynamical stability is demonstrated in terms of the absence of imaginary-frequency phonon modes in \cref{fig:phonons}.
\newline \indent
The magnetization can be understood through the $\pi$ electrons of the fullerenes. 
The $\pi$ system of each molecule features two degenerate orbitals at the Fermi level that transform under the 2D \gls{IR} $E^\prime$ at $\Gamma$. 
These two orbitals are filled with three electrons, with one being unpaired and hence net $1\,\mu_\mathrm{B}$ magnetization per molecule (see \suppmat). 
In the unit cell with two \ce{C26} molecules, these orbitals hybridise into eight bands (four spin-up and four spin-down). 
Filling with six electrons results in four filled spin-up and two filled spin-down bands, leaving two empty spin-down bands. 
The main features of these four bands near the Fermi level are well predicted by a minimum \gls{TB} model (see \suppmat).
\newline \indent 
Embedded into the monolayer structure, exchange interactions between magnetic fullerene molecules facilitate ferromagnetic order, giving rise to spontaneous time\hyp reversal symmetry breaking.
The magnetic ground state is demonstrated in terms of the absence of imaginary magnon modes with a global minimum at $\Gamma$ in \cref{fig:magnons}.
Using the \textsc{Multibinit} package\,\cite{Wojde2013, Evans2014, Eriksson2017}, we determine the magnetic order to persist up to a Curie temperature of $T_\text{C} \sim 25~\text{K}$.
\begin{figure*}[t]
    \centering
    \subfloat[\label{fig:band_struc}]{
      \begin{minipage}[t]{0.38\linewidth}
        \centering
        \includegraphics[width=\linewidth]{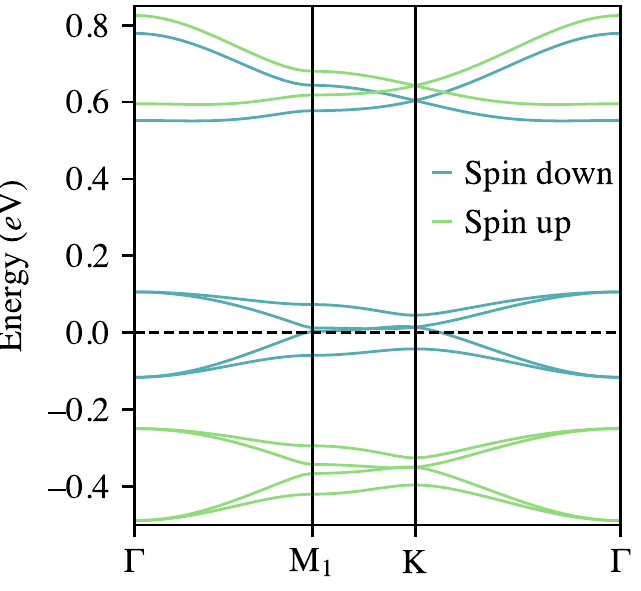}
      \end{minipage}
    }
    \subfloat[\label{fig:weyl_points}]{
      \begin{minipage}[t]{0.61\linewidth}
        \centering
        \includegraphics[width=\linewidth]{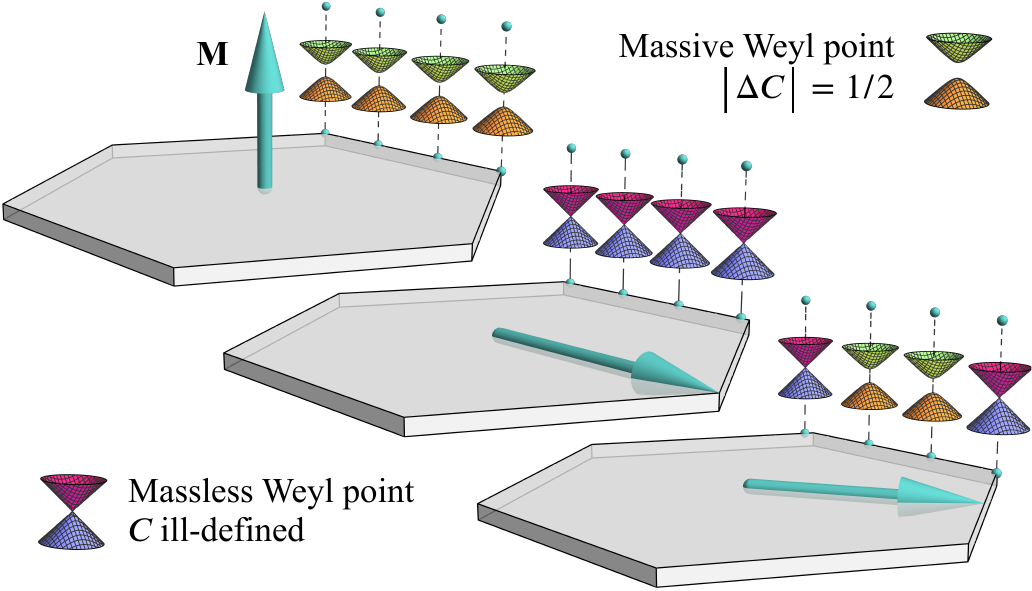}
      \end{minipage}
    }
    \caption{
        (a) Band structure in the absence of \gls{SOC}.
        Bright green (dark blue) bands are spin up (down) polarized.
        The Fermi energy is set to zero and indicated by the dashed line.
        The band structures along any other path $\overline{\Gamma \mathrm{M}_j \mathrm{K}^{(\prime)} \Gamma}$ with $j \in \{1,2,3\}$ is equivalent by symmetry.
        Degeneracies of the two bands in proximity of the Fermi level occur along $\overline{\mathrm{M}_j \mathrm{K}^{(\prime)}}$ (at $\mathrm{K}$) and are due to the two bands transforming under distinct \gls{1D} \glspl{IR} (a \gls{2D} \gls{IR}).
        (b) The hexagonal plane corresponds to the \gls{BZ} and defines $\theta = \pi/2$.
        Weyl points occur on the edges and are exemplified along a $\overline{\mathrm{KK}^\prime}$ path.
        Here, colored surfaces illustrate the dispersion of the two bands close to the Fermi energy around the lattice momentum indicated by the turquoise spheres.
        Upon incorporating \gls{SOC}, the direct band gap between the two bands in proximity of the Fermi energy depends on the orientation of magnetization $\mathbf{M}$. 
    }
    \label{fig:dispersion}
\end{figure*}
The symmetry of the resulting quasi\hyp \gls{2D} crystal structure is described by \gls{SG} $P \overline{6}2m$ (No.\,189).
Our forthcoming discussion focuses on the properties of the two bands in proximity of the Fermi energy in \cref{fig:band_struc}.
In the absence of \gls{SOC}, we determine these states to transform under the \gls{2D} \glspl{IR} of the little co\hyp groups at $\text{K}$ and $\text{K}^\prime$ using the \textsc{IrVsp} software package\,\cite{Gao2021}.
Further two\hyp fold degeneracies are observed along all \gls{BZ} boundaries connecting $\text{M}_{j \in \{1,2,3\}}$ with $\text{K}$ and $\text{K}^{\prime}$.
They are protected by both two\hyp fold rotational symmetry around the axis perpendicular to the respective \gls{BZ} boundary and reflection symmetry about the vertical plane projecting onto this axis.
Consequently, degeneracies between the two relevant bands occur at a total of eight inequivalent lattice momenta.
Locally, we confirm them to disperse linearly, giving rise to fermionic \gls{2D} Weyl quasiparticles with symmetry\hyp protection. 

\begin{figure*}[tb]
    \centering
    \subfloat[\label{fig:phase_diagram_unstrained_analytic}]{
      \begin{minipage}[t]{0.294\linewidth}
        \centering
        \includegraphics[width=\linewidth]{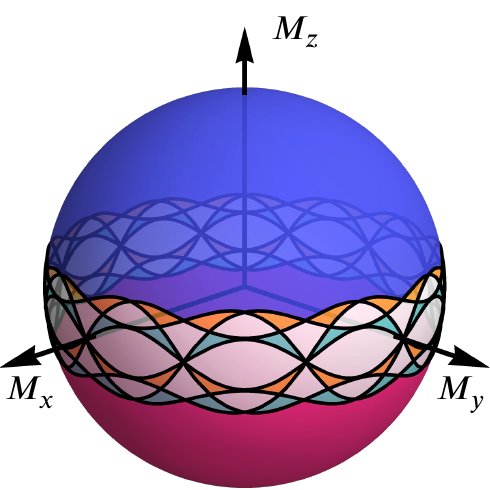}
      \end{minipage}
    }
    \subfloat[\label{fig:phase_diagram_unstrained_numeric}]{
      \begin{minipage}[t]{0.206\linewidth}
        \centering
        \includegraphics[width=\linewidth]{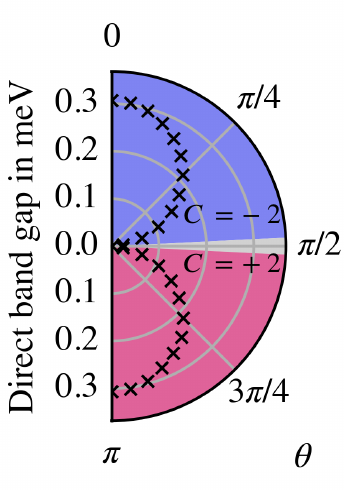}
      \end{minipage}
    }
    \subfloat[\label{fig:phase_diagram_full}]{
      \begin{minipage}[t]{0.5\linewidth}
        \centering
        \includegraphics[width=\linewidth]{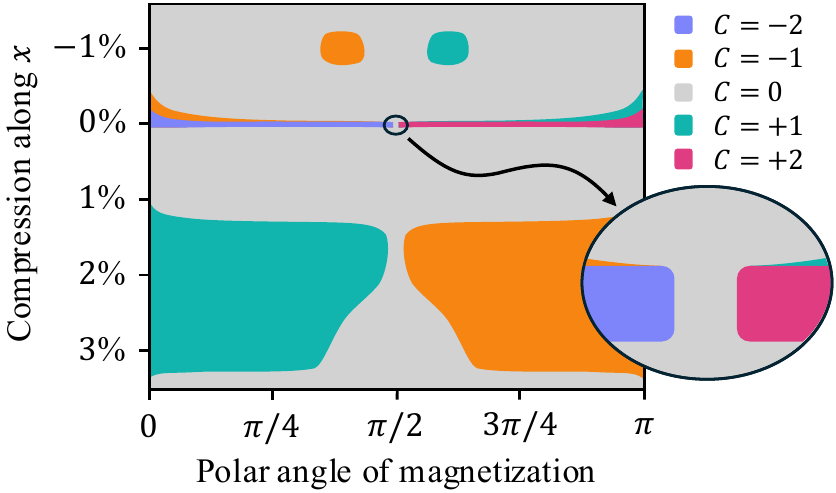}
      \end{minipage}
    }
    \caption{
        (a) Illustration of the theoretically predicted phase diagram.
        Depending on the orientation of magnetization $\mathbf{M}$, different Chern number regimes are identified, indicated by different coloring listed in (c).
        Along black lines, the two bands in proximity of the Fermi energy form massless Weyl points.
        (b) Numerically determined direct band gaps and Chern number regimes for $\phi ~ \text{mod} ~ (\pi/3) = \pi / 6$.
        (c) Phase diagram upon introducing in\hyp plane strain with azimuthal angle $\varphi ~ \text{mod} ~ (2\pi/3) = 2\pi/3$.
        The figure shows the $\phi = \pi / 4$ projection of the parameter space and the details of the interpolation scheme are provided in the \suppmat.
        }
    \label{fig:phase_diagram}
\end{figure*}
Upon introducing \gls{SOC}, the system's ground state energy generally depends on the orientation of magnetization, which is commonly quantified in terms of the \gls{MAE}.
Due to the absence of heavy atoms in our system, however, this effect is negligible, such that magnetic isotropy is maintained up to practically irrelevant corrections (see \suppmat).
Nevertheless, relevant symmetries are now captured by the magnetic double \gls{SG} under which the full Hamiltonian is invariant.
Crucially, in the presence of \gls{SOC}, the system's reflection and rotational symmetries are preserved if and only if the spin magnetization $\mathbf{M} \propto (\sin \theta \cos \phi, \sin \theta \sin \phi, \cos \theta)^\text{T}$ in spherical coordinates is directed perpendicular to the reflection plane and parallel to the rotation axes, respectively.
The dependence is illustrated in \cref{fig:weyl_points}.
Some consequences are in line:
\begin{enumerate*}[label=(\roman*)]
    \item The symmetry\hyp protection of the two Weyl points along the \gls{BZ} boundary on which $\text{M}_{j \in \{1,2,3\}}$ is located becomes broken unless magnetization is in\hyp plane, \ie $\theta = \pi/2$, and $\phi ~ \text{mod} ~ \pi \in \mathcal{M}_j$.
    Here, we define the sets $\mathcal{M}_j \equiv \{ (2j - 1) \pi/6,  (j + 1) \pi/3\}$ such that for $(\theta, \phi ~ \mathrm{mod} ~ \pi) \in \{\pi/2\} \times \mathcal{M}_j$ the direction of magnetization is parallel or perpendicular to the \gls{BZ} boundary on which $\text{M}_{j \in \{1,2,3\}}$ is located.
    The fact that two non\hyp diametral values $\phi \in \mathcal{M}_j$ allow to restore the corresponding Weyl point relates to the latter's protection by both reflection and rotation symmetry.
    \item Three\hyp fold rotational symmetry is preserved only for $\theta \in \{ 0, \pi \}$, where the corresponding little co\hyp groups at $\text{K}$ and $\text{K}^{\prime}$ are reduced to the double group $C_{3h}$ and thus do not feature \gls{2D} \glspl{IR}.
    Consequently, no Weyl points can occur at the \gls{BZ} corner points for any values of $\theta$ and $\phi$ when \gls{SOC} is non\hyp vanishing.
    In the context of a $\mathbf{k} \cdot \mathbf{p}$ model, however, we show that for $\theta = \pi / 2$ and those $\phi$ giving rise to symmetry\hyp protected Weyl points on the \gls{BZ} boundary on which $\mathrm{M}_j$ is located, \gls{SOC} shifts the degeneracies from $\text{K}^{(\prime)}$ onto the same boundary (see \suppmat).
    Any orientation of magnetization $(\theta, \phi) \in \{ \pi / 2 \} \times \mathcal{M}$ with $\mathcal{M} \equiv \cup_j \mathcal{M}_j$ thus exhibits a total of four Weyl points.
    Their protection due to either reflection or two\hyp fold rotational symmetry implies that the bands are fully spin\hyp polarized even in the presence of \gls{SOC}.
    Upon suppressing the electronic density of states at the Fermi energy, \ce{C26} monolayers thus realize a \gls{WHSM}\,\cite{You2019}.
\end{enumerate*}
\newline\indent
As it turns out, these arguments exhaust the symmetry\hyp protected degeneracies between the two relevant bands.
We remark, however, that $\theta$ and $\phi$ provide additional degrees of freedom that can be tuned to close the band gap at some point in the \gls{BZ} even in the presence of only trivial little co\hyp groups.
These so\hyp called \textit{accidental} degeneracies become relevant in the forthcoming discussion.
\textit{Topological properties-- }
Time\hyp reversal symmetry\hyp breaking exhibited by honeycomb \ce{C26} fullerene monolayers originates in the spin\hyp channel.
In the absence of \gls{SOC}, the Berry curvature $\Omega_{xy} (\mathbf{k})$ thus remains odd as a function of lattice momentum $\mathbf{k}$.
Since the Chern number $C \in \mathbb{Z}$ is proportional to the integral of Berry curvature over the \gls{BZ}, topologically non\hyp trivial phases, \ie $C \neq 0$, can emerge only upon coupling spin and spatial degrees of freedom.
Hence, we investigate the system's topological properties with respect to the direct gap between the two electronic bands near the Fermi energy incorporating \gls{SOC}.
Away from the previously discussed orientations of magnetization $(\theta, \phi) \in \{\pi/2\} \times \mathcal{M}$ and $\theta \in \{0, \pi\}$, no symmetry is present that could enforce degeneracies anywhere in the \gls{BZ}.
Weyl points thus generically obtain a finite mass $m$ and contribute $\Delta C = \sgn{m}/2$ to the Chern number.
Any smooth deformation of the system's parameters such that $\Delta C \to - \Delta C$ is consequently accompanied by an intermediate state at which $m=0$ corresponding to a vanishing direct band gap.
\newline \indent
We now examine the implications of two\hyp fold rotational symmetries in $D_{3h}$.
In the presence of \gls{SOC}, these have to be paired with the same rotation of the spin degree of freedom\,\cite{Damhus1984, Liu2022_1, Jiang2024} and thus rotate both the lattice momentum and magnetization while changing the sign of Berry curvature.
From the orientation of rotation axes with respect to Weyl point positions, we conclude that smoothly changing the orientation of magnetization from $\theta = 0$ to $\theta = \pi$ necessitates each Weyl point to change the sign of its mass term via a gap closure.
Vertical reflection and three\hyp fold rotational symmetries further constrain how these gapless phases can emerge in the parameter space spanned by $\theta$ and $\phi$.
For the detailed arguments we refer to the \suppmat.
The resulting phase diagram depending on $\theta$ and $\phi$ is illustrated in \cref{fig:phase_diagram_unstrained_analytic}.
In particular, for in\hyp plane magnetization, \ie $\theta = \pi/2$, we determine a topologically trivial phase which transits into a high Chern number phase with $\abs{C} = 2$ for out\hyp of\hyp plane magnetization.
Honeycomb \ce{C26} monolayers thus exhibit the physics described by Haldane's model\,\cite{Haldane1988}.
\newline \indent
Our theoretical insights are further verified numerically by using \textsc{Wannier90}\,\cite{Pizzi2020} to construct a 
\gls{TB} model representing the isolated group of twelve bands around the Fermi energy shown in \cref{fig:band_struc}.
Topological properties are subsequently deduced using \textsc{WannierTools}\,\cite{Wu2018}, and the resulting phase diagram is exemplified in \cref{fig:phase_diagram_unstrained_numeric}.
In particular, for out\hyp of\hyp plane magnetization, topological phases with $\abs{C} = 2$ are confirmed.
Due to negligible \gls{MAE}, changing the orientation of magnetization using weak external magnetic fields facilitates a simple topological phase\hyp switching mechanism.
\newline \indent
%
    %
    %
        %
        %
%
We further examine the effect of in\hyp plane uniaxial strain.
When applied along any of the high\hyp symmetry directions with azimuthal angle $\varphi \in \mathcal{M}$, one vertical reflection and one two\hyp fold rotational symmetry are preserved, reducing the \gls{SG} to $Amm2$ (No.\,38).
Away from these strain\hyp directions, the symmetry protection of all Weyl points is broken and no further topologically non\hyp trivial phases are expected.
For $\varphi \in \mathcal{M}$ and $(\theta, \phi ~ \text{mod} ~ (\pi/2)) = (\pi/2, \varphi)$, on the other hand, Weyl points along the \gls{BZ} boundary perpendicular or parallel to the axes of strain remain protected.
Hence, under small symmetry\hyp preserving perturbations, they can only be shifted along the lines of mirror and reflection invariant lattice momenta.
We will now focus on the three equivalent directions $\varphi = 2j\pi/3$ with $j \in \{0,1,2\}$ for concreteness and discuss our numerical results.
Depending on the type of structural deformation, we observe two distinct mechanisms: Under tensile strain, Weyl points are shifted away from the $\mathrm{M}_j$ lattice momenta until they meet pairwise in reciprocal space.
Upon further increasing exerted strain, the band degeneracy is lifted.
The pairing is such that driving $(\theta, \phi)$ away from the symmetry\hyp preserving values, always two annihilating Weyl points develop mass terms with opposite signs.
This process is thus not accompanied by a topological phase.
Under compressive strain, on the other hand, Weyl points are shifted towards $\mathrm{M}_j$ and annihilate in pairs of equal mass signs such that two band inversions are observed: For out\hyp of\hyp plane magnetization, the system transits from $C=0$ to $\abs{C} = 1$ for $\sim 1.2 \%$ compression and back to $C=0$ for $\sim 3.4 \%$.
Topological phase boundaries realize Weyl points in the dispersion of the two relevant bands, again giving rise to fine\hyp tuned \gls{WHSM} phases.
These results are visualized in terms of a \gls{2D} slice of the associated phase diagram in \cref{fig:phase_diagram_full}.
\textit{Global Band Gap-- }
At this point, our discussion is still to be taken with a grain of salt: While we have analyzed the direct band gap topology in \ce{C26} monolayers and found the structure to realize a Chern metal\,\cite{Cook2014, Liang2025}, realization of the \gls{QAHE} and fine\hyp tuned emergence of the \gls{WHSM} phase rely on introducing a global band gap and transiting the system into a bulk\hyp insulating state.
For the \ce{C26} monolayer, we find that uniformly exerting tensile strain within the plane levels the energetic gap between valence and conduction band in reciprocal space.
The increased ionic distance reduces the overlap of bonding and anti\hyp bonding atomic orbitals, increasing and decreasing their energy, respectively, which generally yields a reduced band gap\,\cite{Huang2015}.
As moderate uniform strains preserve the symmetry, however, the \gls{SOC}\hyp induced band\hyp repulsion mechanism maintains the direct band gap away from in\hyp plane orientations of magnetization with $\phi \in \mathcal{M}$, causing the Weyl points to localize in energy.
\newline \indent
Since the band gap is opened solely due to \gls{SOC}, the energetic separation of valence and conduction bands in uniformly strained \ce{C26} monolayers is below room temperature energy scale.
Practical application under ambient conditions thus urges heavier ions to contribute to electronic states close to the Fermi energy, which can be accomplished via endohedral doping\,\cite{Zhao2020} or the substrate effect\,\cite{Giovannetti2007}, for example.
We demonstrate the general mechanism by simulating this proximity effect via rescaled \gls{SOC} in the \suppmat.
The resulting direct band gap $\sim \mathcal{O}(10 ~ \mathrm{m}e\mathrm{V})$ gives rise to well\hyp known manifestations of the \gls{QAHE} such as dissipationless conducting edge states under experimentally readily accessible conditions.

\textit{Summary \& Discussion-- }
We demonstrate the superatom ansatz in the context of the \gls{QAHE}.
Taking \gls{2D} honeycomb \ce{C26} fullerene networks as an example, we show how time\hyp reversal symmetry breaking can be achieved by the choice of molecular constituents.
The band topology of the resulting structure is thoroughly investigated, establishing its phase diagram from both theoretical and computational insights.
Easily accessible phase\hyp control mechanisms are identified as the direct band gap increases (i) under moderate strain and (ii) as a consequence of \gls{SOC}.
\newline \indent
While proposals for realizations of the \gls{QAHE} are numerous, most materials are based on complex structures engineered on atomic level, thus evading chemical synthesis capabilities.
Our proposal is solely based on carbon atoms, and the underlying superatom paradigm is anticipated to further facilitate implementation.
While in the absence of heavy atoms only a small band gap emerges, the \ce{C26} monolayer provides a versatile platform to realize various topological phases with several options to tune observation temperature as required.

\textit{Acknowledgments-- }
We are grateful to Prof. Frederick Duncan Michael Haldane at Princeton, Dr Gunnar F. Lange at Oslo, and Mr Wojciech J. Jankowski at Cambridge for helpful discussions and comments.
L.W.P. acknowledges support from the Klaus Höchstetter\hyp Stiftung at Munich and from the Trinity College Cambridge Studienstiftung des deutschen Volkes exchange studentship. J.W. acknowledges support from the Cambridge Undergraduate Research Opportunities Programme and from Peterhouse for the James Porter Scholarship. B.P. acknowledges support from Magdalene College Cambridge for a Nevile Research Fellowship. The calculations were performed using resources provided by the Cambridge Service for Data Driven Discovery (CSD3) operated by the University of Cambridge Research Computing Service (\url{www.csd3.cam.ac.uk}), provided by Dell EMC and Intel using Tier-2 funding from the Engineering and Physical Sciences Research Council (capital grant EP/T022159/1), and DiRAC funding from the Science and Technology Facilities Council (\url{http://www.dirac.ac.uk}), as well as with computational support from the UK Materials and Molecular Modelling Hub, which is partially funded by EPSRC (EP/T022213/1, EP/W032260/1 and EP/P020194/1), for which access was obtained via the UKCP consortium and funded by EPSRC grant ref EP/P022561/1.
\bibliography{Literature}
\end{document}